**Effect of the magnetic moment on the interaction energy between particles.**

Voicu Dolocan

Faculty of Physics, University of Bucharest, Bucharest, Romania

*ABSTRACT*

In this paper we take into account the effect of the magnetic moment on the Yukawa and Coulomb interaction between particles. We find that the Yukawa's and Coulomb's laws are modified by a factor *cosine*, whose argument depends on the magnetic moment of the particles and also, on the distance between the two particles. The obtained results are in a good agreement with experimental data.

## 1. Introduction.

A particle with spin, a particle with charge and orbital angular momentum posses magnetic moments. Each such moment tends to interact with a magnetic field in a quantized manner. Suppose that the wave function for a steady state where there is no charge on the particle is $\phi_o$. The wave function $\phi$ for movement along a given path through a magnetic field is related to the solution $\phi_o$ for the trajectory in absence of the field by the transformation

$$\varphi = \varphi_o \exp\left[\frac{ie}{\hbar c} \int_0^l A \cdot dr \right] \qquad (1)$$

Function $\phi$ in equation (1) then describes the corresponding steady state when the particle carries charge $e$. The unperturbed wave function of the electron is written as a plane wave

$$\varphi_k = \frac{1}{\sqrt{\Omega}} \exp\left[ i\left( k \cdot r + \frac{e}{\hbar c} \int_0^l A \cdot dr - \frac{\varepsilon_k}{\hbar} t \right) \right] \qquad (2)$$

Where $k$ is the wave vector, $A$ is the potential vector, $r$ is the position vector, $\Omega$ is the volume of the system, $\varepsilon_k$ is the energy of the particle. We consider the potential vector

$$A = \frac{\mu \times r}{r^3} \qquad (3)$$

where $\mu$ is the magnetic dipole moment and $r$ is a vector from the middle of the loop to an observation point. The theory and experiment demonstrate that the free electron posses a spin momentum $s$, the projections of which on a specified direction are $s_z = \pm \hbar/2 = \hbar m_s$ where $m_s = \pm \frac{1}{2}$ is the spin quantum number. Also, the electron posses a magnetic moment equal to the Bohr magneton $\mu_B$
We write $\mu_z^{(s)} = \mu_B g m_s$, where $g = 2$. In addition, an electron in a stationary state in an atom, having a definite angular momentum projection $L_z = \hbar m_l$ ($m_l$ is the quantum magnetic number), posses a magnetic moment $\mu_z^{(l)} = \mu_B m_l$, where $\mu_B = e\hbar/2mc$ is the Bohr magneton.

## 2. The Yukawa potential.

It is assumed that an electric charge may be viewed as a source of virtual photons and in the same way the nucleon acts as a source of virtual mesons. First, we present the Yukawa potential and at the end we proceed to the Coulomb potential formulae. We start with the wave equation of a free, relativistic particle with mass $m$. If we replace the energy $E$ and momentum $p - eA/c$ in the energy momentum relationship $E^2 = (p - eA/c)^2 c^2 + m^2 c^4$ by the operators $i\hbar \partial / \partial t$ and $(-i\hbar \nabla + eA/c)$ as it is done in the Schrödinger equation we obtain the Klein-Gordon equation

$$\frac{1}{c^2} \frac{\partial^2}{\partial t^2} \varphi(r,t) = \left[ \left( \nabla + \frac{e}{\hbar c i} A \right)^2 - \frac{1}{\lambda^2} \right] \varphi(r,t) \qquad (4a)$$

where $\lambda = \hbar / mc$ is the Compton wave length of the meson. Suppose that we have a fixed $\delta$-function source for a real scalar field $\phi$, that persists for all time. In order to find $\phi(r)$ we must solve the static Klein-Gordon equation

$$-\left( \nabla + \frac{e}{\hbar c i} A \right)^2 \varphi + \frac{1}{\lambda^2} \varphi = \delta(r) \qquad (4b)$$

We can solve this using the Fourier transform

$$\varphi(r) = \int \frac{dk}{(2\pi)^3} \exp\left[ i\left( k.r + \frac{e}{\hbar c} \oint A.dr \right) \right] \tilde{\varphi}(k)$$

Plugging this into (4b) tell us that $(k^2 + 1/\lambda^2)\varphi(k) = 1$ giving us the solution

$$\varphi(r) = \int \frac{dk}{(2\pi)^3} \frac{\exp\left[ i(k.r) + \frac{e}{\hbar c} \oint A.dr \right]}{k^2 + 1/\lambda^2}$$

Changing to polar coordinates and writing **k.r** = $kr\cos\theta$, we have

$$\varphi(r) = \frac{1}{(2\pi)^2} \int_0^\infty dk \frac{k^2}{k^2 + 1/\lambda^2} \frac{2\sin kr}{kr} \cos\Gamma_o = \frac{1}{(2\pi)^2 r} \cos\Gamma_o \int_{-\infty}^\infty dk \frac{k \sin kr}{k^2 + 1/\lambda^2} =$$

$$\frac{1}{2\pi\pi} \cos\Gamma_o \text{Re}\left[ \int_{-\infty}^\infty \frac{dk}{2\pi\pi} \frac{k\exp(ikr)}{k^2 + 1/\lambda^2} \right]$$

To compute this last integral by closing the contour in the upper half plane $k \to + i\infty$ picking up the pole at $k = +i/\lambda$. This gives

$$\varphi(r) = \frac{1}{4\pi r} \cos\Gamma_o \exp(-r/\lambda) \qquad (5)$$

The field dies off exponentially quickly at distances $\lambda$ and is modulated by a factor cosine. Therefore, the potential oscillates with

$$\Gamma_o = \frac{e}{\hbar c}\oint A.dr = \frac{2\pi e\mu}{\hbar cr}$$

where $\mu$ is the magnetic moment of the nucleon.

Now we consider just two sources at $r_1$ and $r_2$ acting on the vacuum of the boson field. The local number density of these sources (particles) is concentrated like a delta function at each of their centers, $r_j$. The Hamiltonian of the interaction between these sources is written as

$$H_I = g\sum_j \delta(r - r_j)\varphi(r) \qquad (6a)$$

where $\phi(r,t)$ is the solution of the Klein-Gordon equation (4a), which in the second quantization may be written

$$\varphi(r,t) = \frac{1}{\sqrt{V}}\sum_k \sqrt{\frac{\hbar c^2}{2\omega_k}}\left\{a_k \exp\left[i\left(k.r + \frac{e}{\hbar c}\oint A.dr\right) - i\omega_k t\right] + a_k^+ \exp\left[-i\left(k.r + \frac{e}{\hbar c}\oint A.dr\right) + i\omega_k t\right]\right\}$$

$$\omega_k^2 = c^2(k^2 + 1/\lambda^2)$$

$g$ is an arbitrary constant. In the particle representation (6a) becomes

$$H_I = -ig\sum_k \sqrt{\frac{\hbar c^2}{2V\omega_k}}\sum_j \left[a_k^+ \exp\left\{-i\left(k.r_j + \frac{e}{\hbar c}\oint A(r_j)dr\right)\right\} - a_k \exp\left\{i\left(k.r_j + \frac{e}{\hbar c}\oint A(r_j)dr\right)\right\}\right]$$

(6b)

It is easy enough to calculate the effect of such a term as a perturbation. Then the exchange of energy is

$$E_I = \sum_k \frac{|\langle 0|H_I|1_k\rangle|^2}{E(0) - E(k)}$$

Substituting from (6b) we get

$$E_I = \sum_k \frac{1}{-\omega_k}\frac{g^2 c^2}{2\omega_k}\left|\sum_j \exp\left\{i\left(k.r_j + \frac{e}{\hbar c}\oint A(r_j)dr\right)\right\}\right|^2 = -\frac{g^2 c^2}{V}\sum_k \frac{1}{\omega_k^2}[1 + \cos\Gamma]$$

$$\Gamma = k.(r_1 - r_2) + \frac{e}{\hbar c}\left(\oint A_1.dr - \oint A_2.dr\right) = kr\cos\theta - \Gamma_o \qquad (7)$$

$$\Gamma_o = \frac{e}{\hbar c}\left(\oint \mu_1 \times r/r^3 - \oint \mu_2 \times r/r^3\right) = \frac{2\pi\pi}{\hbar cr}(\mu_1 - \mu_2), \quad \mu_1 \uparrow\uparrow \mu_2$$

$$\Gamma_o = \frac{2\pi\pi}{\hbar cr}(\mu_1 + \mu_2), \quad \mu_1 \uparrow\downarrow \mu_2$$

where $r = r_1 - r_2$. The first term in Eq. (7) is

$$E_{I1} = -\frac{g^2}{V}\frac{V}{(2\pi)^3}\int_0^{k_m}\frac{4\pi\pi^2}{k^2+1/\lambda^2}dk = -\frac{g^2}{2\pi^2}\left(k_m - \frac{1}{\lambda}\arctan(\lambda k_m)\right)$$

which evidently tends to infinity with $k_m$, what ever the value $1/\lambda$. For $k_m = \pi/r$, we have

$$E_{I1} = -\frac{g^2}{2\pi r} + \frac{g^2}{2\pi}\frac{1}{\lambda}\arctan(\pi\lambda/r)$$

For $r >> \lambda$, we have $\arctan(\pi\lambda/r) \approx \pi\lambda/r$ and $E_{I1} = 0$. For $r \to 0$, we have $\arctan(\pi\lambda/r) = \pi/2$ and

$$E_{I1} = -\frac{g^2}{2\pi}\left(\frac{1}{r} - \frac{1}{2\lambda}\right) \approx -\frac{g^2}{2\pi r}$$

The second term of Eq. (7) is

$$E_{I2} = -\frac{g^2}{8\pi^3}\int\frac{\cos(kr\cos\theta + \Gamma_o)}{k^2+1/\lambda^2}dk = -\frac{g^2}{4\pi r}\cos(\Gamma_o)\exp(-r/\lambda)$$

Therefore, the interaction energy is

$$E_I = -g^2\left[\frac{1}{2\pi r} + \frac{1}{4\pi r}\cos(\Gamma_o)\exp(-r/\lambda) - \frac{1}{2\pi^2\lambda}\arctan(\pi\lambda/r)\right] \qquad (8)$$

The second term enclosed between parenthesis is modulated by $\cos(\Gamma_o)$; for $\cos(\Gamma_o) = 1$ this is the Yukawa potential[1]. We specify that the classical spin 1 massive particles may be described by the Proca equations[2] for the four vector field. Denoting by $\mu_p$ the magneton of the proton and by $\mu_n$ the magneton of the neutron, we have $\mu_p = 2.7928\mu_N$, $\mu_n = -1.91342\mu_N$, where $\mu_N = e\hbar/2M_p c$, $M_p = 1836.149 m_o$ and $m_o$ is the mass of the free electron. In the case of the proton-proton interaction, $\Gamma_o = 0.085477247\pi/r$, where $r$ is expressed in fm, when the spins are anti parallel, and $\Gamma_o = 0$ when the spins are parallel. In the case of the proton - neutron interaction, if we apply the same expression for $\Gamma_o$, in the case when the magnetic moments are anti parallel we write

$$\Gamma_o = \frac{2\pi\pi}{\hbar c r}(\mu_p + \mu_n) = \frac{0.013436\pi}{r}$$

and when the magnetic moments are parallel,

$$\Gamma_o = \frac{2\pi\pi}{\hbar c r}(\mu_p - \mu_n) = \frac{0.072019.9\pi}{r}$$

where r is expressed in fm. As it appears from Eq. (8) at a distance between nucleons which is larger than the Compton wavelength, $\lambda$, the terms one and three cancels one another and the second term becomes very small.

In the limit $m \to 0$ one obtains the Poisson equation for a space without charges

## 3. Screening effect.

Let us consider that an electron stays at a position and behaviours as an impurity in the electron gas[3]. The unperturbed wave function of the electron is written as a plane wave, $\phi_k(1)$. By assuming a perturbed potential

$$\delta u(r,t) = u \exp\left[i\left(q.r + \frac{e}{\hbar c}\int_0^l \delta A.dr\right)\right]\exp[i\omega t + \alpha t] \quad (9)$$

$\phi_k$ changes into

$$\Psi_k = \varphi_k + b_{k+q}(t)\varphi_{k+q}\exp\left[i(\varepsilon_{k-q} - \varepsilon_k)t/\hbar\right]$$

$$b_{k+q}(t) = \frac{u\exp[i\omega t + \alpha t]}{\varepsilon_k - \varepsilon_{k+q} \pm \hbar\omega \mp i\hbar\alpha} \quad (10)$$

The potential is assumed to oscillate in space and time with Fourier components $q$ and $\omega$, respectively. The time constant $\alpha$ is s positive infinitesimal and $\delta u = 0$ at $t \to -\infty$. The change of charge distribution due to (10) becomes

$$\delta\rho(r,t) = e\sum_{k,\sigma}\left\{|\Psi_k(r,t)|^2 - \frac{1}{\Omega}\right\}f(k)$$

$$\frac{e}{\Omega} + \sum_{k,\sigma}\left\{b_{k+q}(t)\exp\left[i\left(q.r + \frac{e}{\hbar c}\int_0^l \delta A.dr\right)\right] + b_{k+q}(t)\exp\left[-i\left(q.r + \frac{e}{\hbar c}\int_0^l \delta A.dr\right)\right]\right\}f(k) \quad (11)$$

where $f(k)$ is the Fermi distribution function. To make $\delta\rho(r,t)$ a real number we add $\delta u^*$, and obtain

$$\delta\rho = \frac{e}{\Omega}\sum_{k,\sigma}f(k)\left\{\frac{u}{\varepsilon_k - \varepsilon_{k+q} \pm \hbar\omega \mp i\hbar\alpha} + \frac{u}{\varepsilon_k - \varepsilon_{k-q} \mp \hbar\omega \pm i\hbar\alpha}\right\}\times \exp\left[i\left(q.r + \frac{e}{\hbar c}\int_0^l \delta A.dr\right) + i\omega t + \alpha t\right] + C.C.$$

where C. C. means the complex conjugate of the first term. This equation may be rewritten as

$$\delta\rho = \frac{eu}{\Omega}\sum_{k,\sigma}\frac{f(k) - f(k+q)}{\varepsilon_k - \varepsilon_{k+q} + \hbar\omega - i\hbar\alpha}\exp\left[i\left(q.r + \frac{e}{\hbar}\int_0^l \delta A.dr\right) + i\omega t + \alpha t\right] + C.C. \quad (12)$$

The induced charge distribution $\delta\rho$ gives rise to the potential $\delta\Phi(r,t)$ following the modified Poisson equation

$$\left(\nabla + \frac{eA}{\hbar ci}\right)^2 \delta\Phi(r,t) = -\frac{e}{\varepsilon_o}\delta\rho \tag{13}$$

The $\delta\Phi(r,t)$ term is assumed to show the same space and time dependence as $\delta u$

$$\delta\Phi(r,t) = \Phi\exp\left[i\left(q.r + \frac{e}{\hbar c}\int_0^l \delta A.dr\right) + i\omega t + \alpha t\right] + C.C.$$

From (13) $\Phi$ is given by

$$\Phi = \frac{e^2}{\varepsilon_o q^2 \omega}\sum_{k,\sigma}\frac{f(k)-f(k+q)}{\varepsilon_k - \varepsilon_{k+q} + \hbar\omega - i\hbar\alpha}$$

This is the potential created by the redistribution of charge due to $\delta u$ assumed first. This induced potential $\delta\Phi$ is added to the external potential $\delta V(r,t)$ to give rise to $\delta u$ determined self-consistently

$$\delta u(r,t) = \delta V(r,t) + \delta\Phi(r,t)$$

The potential $u$ induced by the external potential $V$ is given by

$$u = \frac{V}{\varepsilon(q,\omega)}; \quad \varepsilon(q,\omega) = 1 + \frac{e^2}{\varepsilon_o q^2 \Omega}\sum_{k,\sigma}\frac{f(k)-f(k\mp q)}{\varepsilon_{k+q} - \varepsilon_k - \hbar\omega + i\hbar\alpha} \tag{14}$$

$\varepsilon(q,\omega)$ is the dielectric function. When the external potential $\delta V(r,t)$ is applied to the electron system, the electrons screen it and

$$\delta u(r,t) = \iint \frac{V(q,\omega)}{\varepsilon(q,\omega)}\exp\left[i\left(q.r + \frac{e}{\hbar c}\int_0^l \delta A.dr\right) + i\omega t\right] dq\,d\omega \tag{15}$$

For the static and smooth spatial change $q \to 0$, so that
$$\varepsilon_{k+q} - \varepsilon_k \approx q.\nabla_k \varepsilon_k$$
$$f(k) - f(k+q) \approx -q\frac{\partial f}{\partial \varepsilon_k}\nabla_k \varepsilon_k$$

and for $\omega = 0$

$$\varepsilon(q,0) \approx 1 + \frac{1}{\lambda_s^2 q^2}; \quad (q \to 0)$$

$$\frac{1}{\lambda_s^2} = \frac{e^2 \rho(\varepsilon_F)}{\varepsilon_o \omega}$$

where

$$\rho(\varepsilon_F) = \frac{\Omega m}{\pi^2 \hbar^3} \sqrt{2m\varepsilon_F}; \quad \varepsilon_F = \frac{\hbar^2 k_F^2}{2m}$$

is the density of state at the Fermi level, and $\lambda_s$ is the screening length. By taking $V(q) = e^2/\varepsilon_o q^2$ and by using the above approximations, equation (15) my be written

$$\delta u(r) = \frac{1}{(2\pi)^3} \int dq \frac{e^2}{\varepsilon_o q^2} \frac{q^2}{q^2 + 1/\lambda_s^2} \exp\left[i\left(q.r + \frac{e}{\hbar c}\int_0^l \delta A.dr\right)\right]$$

To make $\delta u(r)$ a real number we add $\delta u^*$, so that

$$\delta u(r) = \frac{2\pi}{(2\pi)^3} \int_0^\infty dq \int_0^\pi \sin\theta d\theta \frac{e^2}{\varepsilon_o} \frac{q^2}{q^2 + 1/\lambda_s^2} \cos(qr\cos\theta + \Gamma_o) = \frac{e^2}{4\pi\varepsilon_o r} e^{-r/\lambda_s} \cos(\Gamma_o) \quad (16)$$

where $\Gamma_o$ is given by

$$\Gamma_o = \frac{e}{\hbar c}\oint \delta A.dr = \frac{e}{\hbar c^2} \frac{e^2}{4\pi\pi} \frac{h}{e} \oint \frac{(2m_{s2} - 2m_{s1}) \times r}{r^3} dr = \frac{e^2}{2mc^2}\left(\frac{2\pi \times 2m_{s2}}{r} - \frac{2\pi \times 2m_{s1}}{r}\right)$$

$(h/e)m_{s2}$ and $(h/e)m_{s1}$ are the flux vectors. For $m_{s1} = m_{s2} = \frac{1}{2}$, one obtains $\Gamma_o = 0$, that is when the spins of the electrons are oriented in the same direction there is not a modification on the Coulomb's law. At distances $r$ where $\Gamma_o < \pi/2$, the potential energy is only repulsive. For $m = m_o$ (free electron mass) this distance must be larger than 76 nm, while for $m = 10m_o$ it becomes larger than 7.6 nm. $\lambda_s$ is the screening length. For $1/\lambda_s = 0$, absence of the screening, the well known Coulomb potential is obtained.

### 4. Exchange interaction.

Let us consider the Hamiltonian of the Coulomb interaction between electrons

$$H_C = \frac{1}{2}\frac{e^2}{4\pi\varepsilon_o} \iint dr_1 dr_2 \frac{\rho(r_1)\rho(r_2)}{|r_1 - r_2|}$$

where in second-quantization notation $\rho(\mathbf{r})$ is the operator of the electron density[4]

$$\rho(r) = \sum_\sigma \Psi_\sigma^+(r)\Psi_\sigma(r)$$

where $\sigma = \uparrow, \downarrow$ is the spin orientation and $\Psi_\sigma(\mathbf{r})$ is the filed operator, which satisfies the anti commutation relations of the fermions

$$\{\Psi_{\sigma_1}(r_1), \Psi_{\sigma_2}^+(r_2)\} \equiv \Psi_{\sigma_1}(r_1)\Psi_{\sigma_2}^+(r_2) + \Psi_{\sigma_2}^+(r_2)\Psi_{\sigma_1}(r_1) = \delta_{\sigma_1\sigma_2}\delta(r_1 - r_2)$$

$$\{\Psi_{\sigma_1}(r_1), \Psi_{\sigma_2}(r_2)\} = 0; \quad \{\Psi_{\sigma_1}^+(r_1), \Psi_{\sigma_2}^+(r_2)\} = 0$$

Further,

$$H_c = \frac{1}{2}\frac{1}{4\pi\varepsilon_o}\iint dr_1 dr_2 \sum_{\sigma_1,\sigma_2} \Psi_{\sigma_1}^+(r_1)\Psi_{\sigma_1}(r_1)\frac{e^2}{|r_1-r_2|}\Psi_{\sigma_2}^+(r_2)\Psi_{\sigma_2}(r_2) =$$

$$\frac{1}{2}\frac{1}{4\pi\varepsilon_o}\iint dr_1 dr_2 \sum_{\sigma_1,\sigma_2} \Psi_{\sigma_1}^+(r_1)\Psi_{\sigma_2}^+(r_2)\frac{e^2}{|r_1-r_2|}\Psi_{\sigma_2}(r_2)\Psi_{\sigma_1}(r_1) + \frac{1}{2}\frac{1}{4\pi\varepsilon_o}\int dr_1 \sum_{\sigma_1} \frac{e^2}{|r_1-r_2|}\Psi_{\sigma_1}^+(r_1)\Psi_{\sigma_1}(r_1)$$
(17)

The last term is unphysical. We expand $\Psi$ into a set of orthonormal functions $\varphi_{m_l}(r - R_j)$ localized at the ionic position $\mathbf{R}_j$ (Wannier functions) and a set of spinors

$$\chi_\uparrow \begin{pmatrix} 1 \\ 0 \end{pmatrix}, \chi_\downarrow \begin{pmatrix} 0 \\ 1 \end{pmatrix}$$

which are eigenvectors of

$$s^z = \frac{1}{2}\sigma^z = \frac{1}{2}\begin{Bmatrix} 1 & 0 \\ 0 & -1 \end{Bmatrix}$$

to eigenvalues ± ½. So,

$$\Psi_\sigma(r) = \sum_{j,m_l} a_{jm_l\sigma}\varphi_{m_l}(r - R_j)\chi_\sigma \qquad (18)$$

where $a_{jm_l\sigma}$ is a fermion annihilation operator satisfying $\{a_{jm_l\sigma}, a_{j'm_l'\sigma'}^+\} = \delta_{jj'}\delta_{m_lm_l'}\delta_{\sigma\sigma'}$, etc.

Here $m_l$ includes all orbital quantum numbers but not the spin. The scalar products of the spinors are simple: $\chi_{\sigma_1}^+\chi_{\sigma_1} = \chi_{\sigma_2}^+\chi_{\sigma_2} = 1$. On-site Hamiltonian coulomb interaction is

$$H_c \approx \frac{1}{2}\sum_j \sum_{m_1,m_2} \sum_{\sigma_1,\sigma_2} \{K_{m_1,m_2} a_{jm_1\sigma_1}^+ a_{jm_1\sigma_1} a_{jm_2\sigma_2}^+ a_{jm_2\sigma_2} - J_{m_1,m_2} a_{jm_1\sigma_1}^+ a_{jm_1\sigma_2} a_{jm_2\sigma_2}^+ a_{jm_2\sigma_1}\} +$$

(irrelevant potential term)
(19)

where

$$K_{m_1 m_2} = \iint dr_1 dr_2 |\varphi_{m_1}(r_1 - R_j)|^2 \frac{e^2}{4\pi\varepsilon_o |r_1 - r_2|} |\varphi_{m_2}(r_2 - R_j)|^2 \quad (20)$$

is the direct Coulomb integral, and

$$J_{m_1 m_2} = \iint dr_1 dr_2 \varphi_{m_1}(r_1 - R_j) \varphi_{m_2}(r_2 - R_j) \frac{e^2}{4\pi\varepsilon_o |r_1 - r_2|} \varphi_{m_1}(r_2 - R_j) \varphi_{m_2}(r_1 - R_j) \quad (21)$$

is the exchange integral[5]. Now, instead of

$$\frac{1}{4\pi} \frac{1}{|r_1 - r_2|} = \frac{1}{(2\pi)^3} \int \frac{dk}{k^2} \exp[ik(r_1 - r_2)]$$

we write

$$\frac{1}{(2\pi)^3} \int \frac{dk}{k^2} \exp\left[ik(r_1 - r_2) + \frac{ie}{\hbar c}\left(\oint A(r_1)dr - \oint A(r_2)dr\right)\right] = \frac{1}{4\pi} \frac{\cos(\Gamma_o)}{|r_1 - r_2|} \quad (22)$$

so that in the above expressions instead of $e^2$ we write $e^2 \cos(\Gamma_o)$, where

$$\Gamma_o = \frac{e}{\hbar} \frac{e^2}{4\pi\pi c^2} \frac{h}{e} \oint \frac{(m_{l1} - m_{l2}) \times r}{r^3} = \frac{\pi e^2}{mc^2} \frac{m_{l1} - m_{l2}}{r} \quad (23)$$

where $m_l$ is the magnetic orbital quantum number, $r = |\mathbf{r}_1 - \mathbf{r}_2|$ and $m$ is the effective mass of the electron. The sign minus is when the magnetic moments of the two electrons are oriented in the same direction (ferromagnetic) and the sign plus is then when the magnetic moments of the two electrons are anti parallel (anti ferromagnetic). For $m_{l1} = 2$ and $m_{l2} = 1$, we have $\Gamma_o^F = \pi e^2 / mrc^2$ for parallel orientation and $\Gamma_o^{AF} = 3\pi e^2 / mrc^2$ for the anti parallel orientation of the magnetic moments of the two interacting electrons. In order to introduce into Hamiltonian the product of the spin vectors we proceed as follows[4]. We define the spin operators

$$s_{jm}^\alpha = \sum a_{jm\sigma} \cdot (\sigma_{\sigma \cdot \sigma'} / 2) a_{jm\sigma'}$$

where summation is performed over σ,σ'.
and the number operators $n_{jm} = \sum_\sigma a^+_{jm\sigma} a_{jm\sigma}$ so that (21) becomes

$$H_C = \frac{1}{2} \sum_j \sum_{m_1 m_2} \left\{ (K_{m_1 m_2} - \frac{1}{2} J_{m_1 m_2}) n_{jm_1} n_{jm_2} - 2 J_{m_1 m_2} s_{jm_1} \cdot s_{jm_2} \right\} \quad (24)$$

The first term is the on-site Coulomb interaction. May be shown that $K_{m_1 m_2} - (1/2) J_{m_1 m_2} > 0$, so that the Coulomb term is reduced but still repulsive. The second term $-J_{m_1 m_2} s_{jm_1} . s_{jm_2}$, with $J_{m1m2} > 0$, is the spin-spin interaction contribution. The minimum of the energy acquires the parallel spins alignment, i.e., ferromagnetic. This is the derivation of Hund's first rule. For a single relevant orbital $\phi(r)$, $H_C$ (17) becomes Hubbard[6] $U$ term

$$H_C = \frac{1}{2}\sum_j \iint dr_1 dr_2 \varphi(r_1-R_j)\varphi(r_2-R_j)\frac{e^2\cos(\Gamma_o)}{4\pi\varepsilon_o|r_1-r_2|}\varphi(r_2-R_j)\varphi(r_1-R_j)\sum_{\sigma_1\sigma_2} a^+_{j\sigma_1}a^+_{j\sigma_2}a_{j\sigma_2}a_{j\sigma_1} =$$

$$\frac{1}{2}\sum_j \iint dr_1 dr_2 \varphi(r_1-R_j)\varphi(r_2-R_j)\frac{e^2\cos(\Gamma_o)}{4\pi\varepsilon_o|r_1-r_2|}\varphi(r_2-R_j)\varphi(r_1-R_j)(a^+_{j\uparrow}a^+_{j\downarrow}a_{j\downarrow}a_{j\uparrow} + a^+_{j\downarrow}a^+_{j\uparrow}a_{j\uparrow}a_{j\downarrow}) =$$

$$\sum_j \iint dr_1 dr_2 \varphi(r_1-R_j)\varphi(r_2-R_j)\frac{e^2\cos(\Gamma_o)}{4\pi\varepsilon_o|r_1-r_2|}\varphi(r_2-R_j)\varphi(r_1-R_j)a^+_{j\uparrow}a^+_{j\downarrow}a_{j\downarrow}a_{j\uparrow} =$$

$$\sum_j U a^+_{j\uparrow}a^+_{j\downarrow}a_{j\downarrow}a_{j\uparrow}$$

(25)

where $\cos(\Gamma_o) = 1$, because $m_{l1} = m_{l2}$. Now, we consider inter-ion exchange interaction. The Coulomb Hamiltionian of interaction between electrons $i$ and $j$ belonging to ions positioned in the positions $R_i$ and $R_j$ is

$$H_C = \frac{1}{2}\sum_{ij}\left\{(K_{ij} - \frac{1}{2}J_{ij})n_i n_j - 2J_{ij} s_i . s_j\right\}$$

where

$$K_{ij} = \iint dr_i dr_j |\varphi(r_i-R_i)|^2 \frac{e^2\cos(\Gamma_o)}{4\pi\varepsilon_o|r_i-r_j|}|\varphi(r_j-R_j)|^2$$

$$J_{ij} = \iint dr_i dr_j \varphi(r_i-R_i)\varphi(r_j-R_j)\frac{e^2\cos(\Gamma_o)}{4\pi\varepsilon_o|r_i-R_j|}\varphi(r_j-R_i)\varphi(r_i-R_j)$$

(26)

The Hamiltonian of exchange interaction is[7]

$$H_{exc} = -\sum_{ij} J_{ij} s_i . s_j$$

(27)

We assume that there is a slow increase of the wide of the $d$-band with the increase of the electrons number in it. The narrow is the band, the large is the effective mass of the electrons. So, the effective mass of electron in 3$d$-band of $Cr$ ($3d^5$) is some larger than the effective mass of electron in $Fe$ ($3d^6$), $Co$ ($3d^7$), $Ni$ ($3d^8$). The difference between the values of the distance to the nearest neighbours in these four metals is lower than 0.1%.: Cr ($r_o = 0.2494$ nm,$bcc$; 0.26 nm,$fcc$), Fe (0.2486 nm,$bcc$), Co (0.251 nm,$fcc$), Ni (0.24886 nm,$fcc$). In ferromagnets must be $\cos(\Gamma_o^F) = 1$ and $J_{ij}$ (28) $>0$, while in antiferromagnets we must have $\cos(\Gamma_o^{AF}) = 1$ and $J_{ij} < 0$.

**5. Conclusions.**

We have studied the effect of the magnetic moment of a particle on the energy of interaction between particles. We have found that both Yukawa and Coulomb potentials are modified by a factor on the form of *cosine* whose argument depends on the both orbital and spin magnetic moments. The obtained results are in a good agreement with experimental data[8].